\documentclass[prl,aps,showpacs,epsf,superscriptaddress,twocolumn]{revtex4}

\def\be{\begin{equation}}
\def\ee{\end{equation}}

\usepackage{bm}
\usepackage{amsmath,epsfig}
\usepackage{color}
\usepackage[usenames,dvipsnames]{xcolor}
\usepackage{mathrsfs}
\usepackage{soul} 
\usepackage{float} 

\newcommand{\fgies}[3]{\mbox{\raisebox{#3} 
{\epsfig{file=#1,scale=#2,clip=true}}~}}
\newcommand{\bea}{\begin{eqnarray}}
\newcommand{\eea}{\end{eqnarray}}
\newcommand{\bi}{\begin{itemize}}
\newcommand{\ei}{\end{itemize}}
\newcommand{\smeq}{\,{=}\,}

\newcommand{\rr}{{\bf r}}
\newcommand{\kk}{{\bf k}}

\newcommand{\vn}{{\bf 0}}

\newcommand{\up}{\uparrow}
\newcommand{\down}{\downarrow}

\newcommand{\Cr}{\mathcal{C}}

\begin{document}

\title{Contact and Momentum Distribution of the Unitary Fermi Gas}

\author{R. Rossi}
\altaffiliation{Present address: Center for Computational Quantum Physics,
The Flatiron Institute, New York, USA.}
\affiliation{Laboratoire de Physique Statistique, Ecole Normale Sup\'erieure, Universit\'e PSL, Sorbonne Universit\'e, Universit\'e Paris Diderot, CNRS, Paris, France}
\author{T. Ohgoe}
\affiliation{Department of Applied Physics, University of Tokyo,
7-3-1 Hongo, Bunkyo-ku, Tokyo 113-8656, Japan}
\author{E. Kozik}
\affiliation{Physics Department, King's College,
London WC2R 2LS, United Kingdom}
\author{N. Prokof'ev}
\affiliation{Department of Physics, University of Massachusetts, Amherst, MA 01003, USA}
\affiliation{National Research Center ``Kurchatov Institute", 123182 Moscow, Russia}
\author{B. Svistunov}
\affiliation{Department of Physics, University of Massachusetts, Amherst, MA 01003, USA}
\affiliation{National Research Center ``Kurchatov Institute", 123182 Moscow, Russia}
\affiliation{Wilczek Quantum Center, School of Physics and Astronomy and T. D. Lee Institute, Shanghai Jiao Tong University, Shanghai 200240, China}
\author{K. Van Houcke}
\affiliation{Laboratoire de Physique Statistique, Ecole Normale Sup\'erieure, Universit\'e PSL, Sorbonne Universit\'e, Universit\'e Paris Diderot, CNRS,  Paris, France}
\author{F. Werner}
\affiliation{Laboratoire Kastler Brossel, Ecole Normale Sup\'erieure, Universit\'e PSL, CNRS, Sorbonne Universit\'e, Coll\`ege de France, Paris, France}

\date{\today}

\begin{abstract}
A key quantity 
in strongly-interacting resonant Fermi gases
is the contact $\Cr$,
which characterizes numerous properties such as the momentum distribution at large momenta
or the pair correlation function at short distances.
The temperature dependence of $\Cr$ 
 was 
 measured at unitarity,
 where
 existing theoretical predictions
differ substantially even at the qualitative level.
 We report 
  accurate data
 for the contact 
and the momentum distribution 
of the unitary gas  in the normal phase,
obtained by
Bold Diagrammatic Monte Carlo
and Borel resummation. 
Our results agree with
experimental data within
 error bars and provide crucial benchmarks for the development of advanced
theoretical treatments and precision measurements.

\end{abstract}
\maketitle

The resonant  Fermi gas is a 
fundamental model of quantum many-body physics.
It features a smooth crossover between fermionic and bosonic superfluidity,
as predicted in the context of condensed matter physics ~\cite{Leggett1980_1,Leggett1980_2,NSR,ChapLeggettBref}
and confirmed by 
remarkable
experiments on ultracold atomic Fermi gases
near Feshbach resonances~\cite{ZwergerBook}.
It is also relevant to neutron matter~\cite{GezerlisNeutronsReview,StrinatiUrbanReview} and high-energy physics~\cite{NJP_QCD},
particularly
in the central region of 
the crossover, around the unitary point
where the scattering length diverges.
As a result of
the vanishing interaction range,
resonant  Fermi gases feature characteristic ultraviolet singularities 
 governed by a single quantity called contact~\cite{TanEnergetics,TanLargeMomentum,ChapLeggettBref,ChapBraatenBref,LeChapitreIn}.
In particular,
for the homogeneous gas,
the
density-density correlation function
at
short distance diverges as 
\be
\langle \, \hat{n}_\up(\rr) \,
\hat{n}_\down(\vn) \,
 \rangle
\underset{r\to0}{\sim}\frac{\Cr}{(4\pi\, r)^2}
\label{eq:C_g2}
\ee
and the
momentum distribution has the 
tail 
\be
n_\sigma(\kk)\underset{k\to\infty}{\sim}\frac{\Cr}{k^4}.
\label{eq:C_nk}
\ee
Here
$\Cr$ is the contact per unit volume,
$\hat{n}_\sigma(\rr)=\hat{\psi}^\dagger_\sigma(\rr)\hat{\psi}_\sigma(\rr)$
is the density operator,
and the spin-$\sigma$ momentum distribution $n_\sigma(\kk)$ is 
normalised to $\int n_\sigma(\kk) d^3k / (2\pi)^{3}=n_\sigma = \langle\hat{n}_\sigma(\rr)\rangle$.
A direct manifestation of Eq.~(\ref{eq:C_g2}) is that
in a unit volume,
the number  of pairs of fermions 
separated by a distance smaller than $s$ is
$\Cr\,s / (4\pi)$ in the $s{\to}0$ limit.
Hence $\Cr$ 
controls the (anomalously high) density of pairs with vanishing interparticle distance~\cite{TanEnergetics,BraatenC,ChapBraatenBref}.


A large variety of experimentally studied observables
are directly expressible in terms of the contact:
the population of the closed channel molecular state measured by laser molecular spectroscopy~\cite{HuletClosedChannel,WTC},
the large-momentum tail of the static structure factor measured by Bragg spectroscopy~\cite{AustraliensC, AustraliensT,ValeC_precise},
the tail of the
 momentum distribution 
 measured by non-interacting time-of-flight
 or by momentum-resolved radiofrequency spectroscopy~\cite{JinUnivRel},
the derivative of the energy with respect to the inverse scattering length~\cite{TanLargeMomentum}
extracted from the pressure equation of state measured by in-situ imaging~\cite{NirEOS},
the large-frequency tail in radiofrequency spectroscopy~\cite{RanderiaRF,JinUnivRel,Jin_C_homogeneous,SagiContactRF},
 and
the short-distance density-density correlation function 
extracted from the three-body loss rate in presence of a 
bosonic cloud~\cite{LaurentC}.

The 
experimental study~\cite{Jin_C_homogeneous} is particularly important because
it is spatially resolved and for the first time yields the temperature dependence of the contact
for a homogeneous system. 
Recently, 
two other experimental groups have presented preliminary data for the temperature-dependent homogeneous contact~\cite{ContactSantFeliu}.
Understanding the experimental data remains a major challenge,
because existing theoretical predictions,
based on lowest order skeleton Feynman diagrams~\cite{StrinatiC,ZwergerViscosity,Hu_C}
or 
Monte Carlo simulations on a lattice~\cite{Drut_C,Goulko_UFG_2016}
contradict each other even at the qualitative level, especially on approach to the
superfluid transition from the normal side.



In this Letter, we present 
high precision
results
for the contact 
of the unitary Fermi gas in the normal phase. 
We employ the  Bold Diagrammatic Monte Carlo (BDMC) technique,
in which all skeleton Feynman
diagrams 
are sampled stochastically up to a maximal order $N_{\rm max}$~\cite{VanHouckeEOS},
and convergence towards the exact result in the 
limit $N_{\rm max} \to \infty$  is obtained by applying an appropriate conformal-Borel resummation to the divergent diagrammatic series~\cite{RossiEOS}.
Our results agree with experimental data within the uncertainty limits, and establish that
 the
contact 
is a slowly decreasing  function of temperature
at fixed density in the normal phase.
Furthermore we 
observe
 a non-Fermi liquid behavior
in
the momentum distribution.


We  directly extract the contact from the pair propagator $\Gamma$
thanks to the relation
\be
\Cr = - 
\Gamma(\rr=\vn,\tau=0^-)
\label{eq:CvsGamma}
\ee
(we set $\hbar$ and $m$ to unity).
While
this relation
was first obtained within 
 $T$-matrix approximations
\cite{StrinatiRF,ZwergerRFLong,Hu_C},
it actually becomes exact
once $\Gamma$ is fully dressed.
Physically,
this  relation 
is consistent with the 
interpretation of $\Cr$ in terms of a density of pairs,
since it
is formally analogous to the relation
$n_\sigma\,{=}\,G_\sigma(\rr\,{=}\,\vn,\tau\smeq0^-)$
 between the single-particle density 
 and the single-particle propagator $G$.
A simple way to derive Eq.~(\ref{eq:CvsGamma})
is to use the regularized version of Eq.~(\ref{eq:C_g2})
which
holds in a lattice model~\cite{WernerCastinRelationsFermions},
\be
\Cr
=
g_0^{\phantom{0}2}\,
\langle \hat{n}_\up(\vn) \,
\hat{n}_\down(\vn) 
 \rangle
\label{eq:C_g2_latt}
\ee
i.e. the contact is equal to the double occupancy,
up to a renormalization factor set by the bare coupling constant $g_0$
(see also~\cite{BraatenC}).
The result (\ref{eq:CvsGamma}) then follows from the fact that
\be
\Gamma(\rr,\tau) = g_0 \, \delta(\tau) \frac{\delta_{\rr,\vn}}{b^3}
- g_0^{\phantom{0}2}
\,
\big\langle 
{\rm T}
(\psi_\down \psi_\up)(\rr,\tau)
(\psi^\dagger_\up \psi^\dagger_\down)(\vn,0)
\big\rangle
\ee
or, diagrammatically,
\be
{\hskip -2cm} \fgies{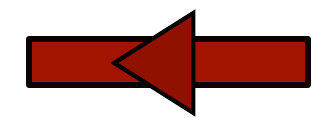}{0.23}{-1.3mm} = 
\,  \fgies{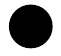}{0.23}{-.5mm}
 \, + \,
   \fgies{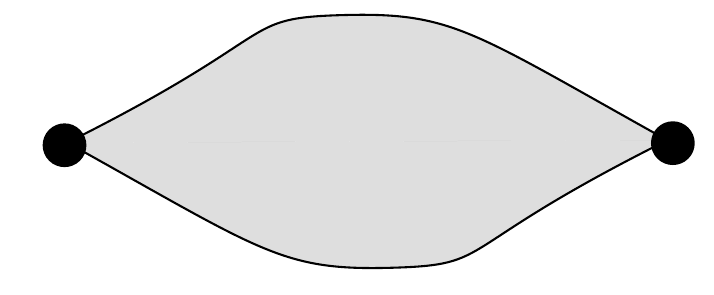}{0.23}{-5mm}.
\ee
Here ${\rm T}[\ldots]$ is the time-ordered product,
and
the first term does not contribute in the continuum limit
where the lattice spacing $b$ 
tends to zero.




Most results presented below were obtained using the bold  scheme, where 
diagrams are built self-consistently from fully dressed propagators $G$ and $\Gamma$;
when the temperature is not too low, we can alternatively use
  the non self-consistent ladder scheme, where diagrams are built from the non-interacting $G_0$ and ladder-sum $\Gamma_0$~\cite{RossiEOS,BDMC_long,ResonLong2}.
By scale invariance, $\mathcal{C}\lambda^4$ is a universal function of $\beta\mu$, with 
$\mu$ the chemical potential, $\beta=(k_B T)^{-1}$ the inverse temperature, and
$\lambda=\sqrt{2\pi\beta}$ the thermal wavelength.
We have cross-checked the bold scheme against the ladder scheme at $\beta\mu=0$, finding a relative difference for $\mathcal{C}\lambda^4$ smaller than
$10^{-4}$,
well within the error bars.
We typically went up to diagram order $N_{\rm max}= 9$~\footnote{See Supplemental Material for our data in numerical form, as well as  plots showing the dependence on $N_{\rm max}$ and a more detailed comparison with the virial expansion.}.
Without resummation, the bold scheme with $N_{\rm max}=1$ coincides with the self-consistent $T$-matrix approximation of Refs.~\cite{Haussmann_PRB,ZwergerViscosity}, and the ladder-scheme with $N_{\rm max}=0$ coincides with the non-self-consistent $T$-matrix approximation of Ref.~\cite{StrinatiC}. The non-resummed results oscillate wildly as a function of $N_{\rm max}$, which illustrates the absence of a small expansion parameter -- {\it e.~g.}, at $\beta\mu=0$ for the ladder scheme, $\Cr\lambda^4$ changes by a factor $\approx 2$ between $N_{\rm max} = 5$ and 6.
We went down to $T/T_F\approx 0.19$, about 10\% above the transition temperature to the superfluid phase $T_c/T_F\approx 0.17$~\cite{KuEOS,GoulkoBurovski}.
Approaching closer to $T_c$ requires tricks to stabilize the bold self-consistency loop, which we leave for future work.

\begin{figure}[h!]
\includegraphics[width=\columnwidth,trim={0cm 0.5cm 1cm 0.5cm},clip=]{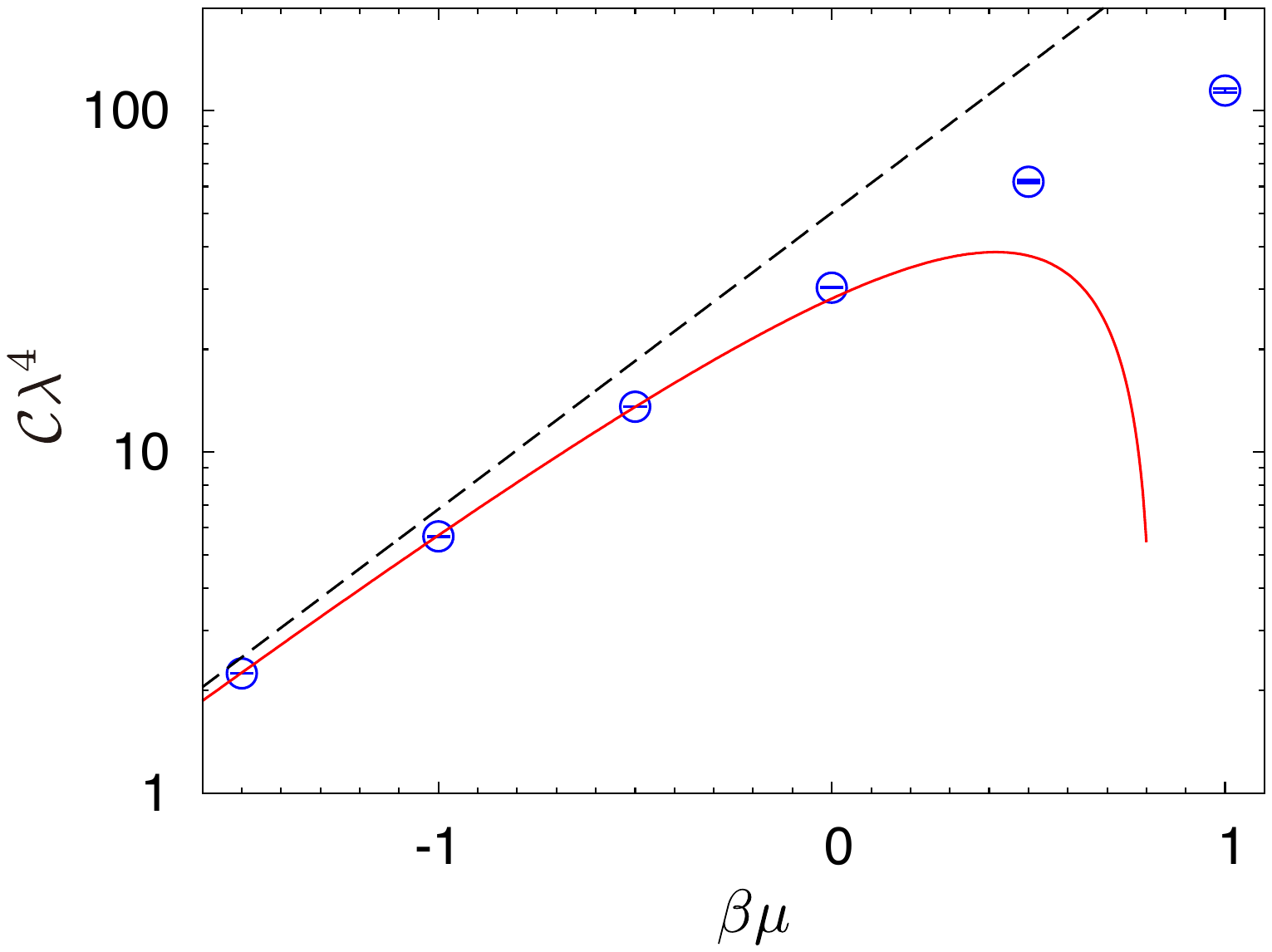}
\caption{The contact obtained
by diagrammatic Monte Carlo (circles with error bars) 
agrees with 
the virial expansion~\cite{Baym_C,SunVirial3} at order two (dashed line) and three (solid line) in the high-temperature limit $\beta\mu\to-\infty$.
\label{fig:Cvirial}}
\end{figure}

Our results in the high-temperature region are shown in Fig.~\ref{fig:Cvirial},
together
with the virial expansion
\be
\Cr \lambda^4 = 16 \pi^2 \left( c_2 \, e^{2\beta\mu} + c_3 \, e^{3\beta\mu} + \ldots \right)
\ee
The coefficients
$c_2=1/\pi$~\cite{Baym_C} 
and 
$c_3 = -0.1399(1)$~\cite{SunVirial3,LeyronasPrivate}\footnote{
See also Ref.~\cite{Hu_C} which finds $c_3\simeq-0.141$.}
come from the $2$-body and $3$-body problem respectively.

\begin{figure*} 
\begin{center}
\includegraphics[width=2\columnwidth,trim={1cm 1cm 1cm 2cm}]{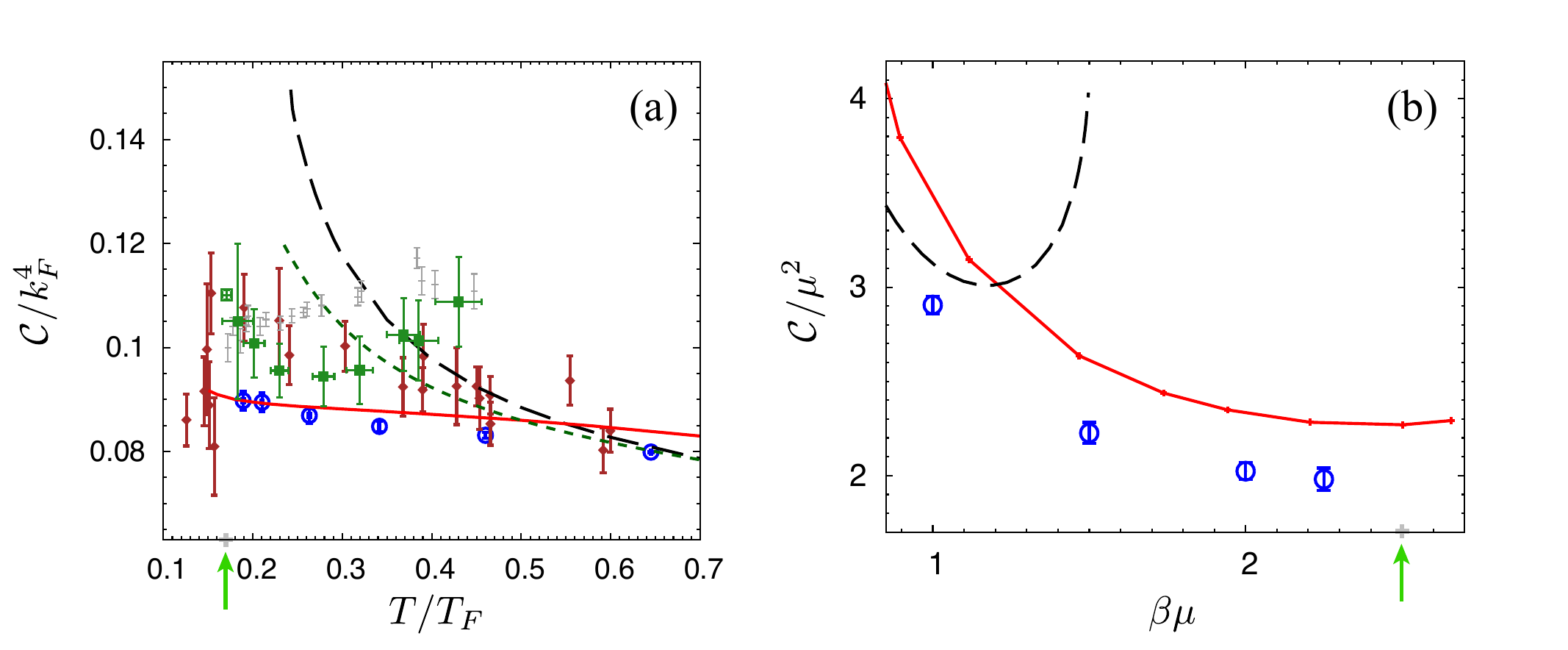}
\caption[width= 2\columnwidth]{Temperature dependence of the contact in the low-temperature region of the normal phase, in terms of (a) canonical
and (b) grand canonical variables.
BDMC (this work): blue solid circles, 
JILA experiment~\cite{Jin_C_homogeneous}: brown solid diamonds,
lattice AFQMC simulations~\cite{Drut_C}: grey crosses,
lattice DDMC with continuous-space and thermodynamic-limit extrapolations~\cite{Goulko_UFG_2016,GoulkoProceeding2010}: green squares.
The curves correspond to different diagrammatic approximations: 
non-self-consistent $T$-matrix~\cite{StrinatiC}: dashed black line, 
self-consistent $T$-matrix~\cite{ZwergerViscosity}: solid red line, 
Nozi\`eres-Schmitt-Rink~\cite{Hu_C}: dotted green line in (a) and dashed black line in (b).
The transition point to the superfluid phase~\cite{KuEOS,GoulkoBurovski}
is  indicated by the green arrow.
\label{fig:C}}
\end{center}
\end{figure*}

The
behavior of the contact in the
low-temperature region of the normal phase is displayed in Fig.~\ref{fig:C}.
The contact in terms of canonical variables, $\Cr(n,T)$
or equivalently $\Cr/k_F^{\phantom{F}4}$ {\it versus} $T/T_F$,
is shown in Fig.~\ref{fig:C}a.
We find a remarkably weak temperature dependence, which results from a compensation between two competing effects, as we will see from the momentum distribution below.
The difference between the experimental results of Ref.~\cite{Jin_C_homogeneous} and our data is on the order of the experimental error bars,
and the sign of this difference is essentially constant which indicates that the experimental error is mostly systematic rather than statistical.
The
lattice Auxiliary-Field Quantum Monte Carlo (AFQMC) data of Ref.~\cite{Drut_C}
disagree with our results and
predict an opposite temperature dependence;
this may be due to a lack of control over
systematic errors, whose main source
is believed to be the
discretization of space (i.e. the finite filling factor)~\cite{Drut_C}.
The Determinantal Diagrammatic Monte Carlo (DDMC) data of Refs.~\cite{Goulko_UFG_2016,GoulkoProceeding2010}
have a
 non-monotonic temperature dependence which 
may also be
 an artifact of space-discretization errors, even though continuous-space extrapolation was performed.
The data point at $T_c$
 from Ref.~\cite{GoulkoProceeding2010} (open square in Fig.~\ref{fig:C}) in combination with our data indicate that the slope $|d(\Cr/k_F^4)/d(T/T_F)|$, which is much smaller than unity in the region $0.19 \lesssim T/T_F \lesssim 1$, quickly increases to values $\gtrsim 1$ on approach to the critical temperature~\footnote{In Ref.~\cite{GoulkoProceeding2010}, the result $\Cr/k_F^4=0.110(1)$  at $T_c$ was obtained by a linear extrapolation of $\Cr/k_F^4$ {\it vs.} $\nu^{1/3}$ where $\nu$ is the filling factor. Using a quadratic extrapolation gives 0.105(5) which leads to the same qualitative conclusion.}
This 
change of behavior may be related to the critical behavior~\footnote{The leading singular part is $\Cr_{\rm sing}(T) \sim \pm A_{\pm}|T-T_c|^{1-\alpha}$ for $T\to T_c^\pm$ with $\alpha=-0.0151(3)$ and $A_+/A_- = 1.061(2)$ for the $U(1)$ universality class~\cite{hasenbusch_exponents,hasenbusch_amplitude_ratio}.}.
The non-self-consistent $T$-matrix results of  Ref.~\cite{StrinatiC}
 and the Nozi\`eres Schmitt-Rink results of  Ref.~\cite{Hu_C}
predict a more pronounced and gradual enhancement of the contact
when decreasing temperature,
which was interpreted in Ref.~\cite{StrinatiC} as a manifestation of pseudogap physics.
Our data demonstrate that this
behavior
is an artifact of the non-self-consistent $T$-matrix approach.
The self-consistent $T$-matrix results of  Ref.~\cite{ZwergerViscosity}
are 
 remarkably close to our data in Fig.~\ref{fig:C}a.
 In Fig.~\ref{fig:C}b we show
 the contact in terms of grand canonical variables,
 $\Cr(\mu,T)$ or equivalently $\Cr/\mu^2$ {\it versus} $\beta\mu$.
 It is natural to use these variables
 to discuss
the different diagrammatic results
 since
the diagrammatic technique is formulated in the grand-canonical ensemble.
In this sense, the function $\Cr(n,T)$ 
is a combination of $\Cr(\mu,T)$ and of the equation of state $n(\mu,T)$,
given for each of the considered approaches in Refs.~\cite{RossiEOS,ChapStrinatiBref,HaussmannZwergerThermo,Hu_PRA_2008}.
The
non-self-consistent $T$-matrix and Nozi\`eres-Schmitt-Rink approaches yield the same result for $\Cr(\mu,T)$
(derivable from Eq.~(\ref{eq:CvsGamma})
by replacing the exact
$\Gamma$ with 
the sum 
 of the ladder diagrams
built on ideal-gas propagators),
featuring again a strong enhancement at low temperature in disagreement with our results.
The self-consistent $T$-matrix  data
follows the same trend as ours
up to a difference of about $20\%$.
This difference
 largely cancels out 
 with the difference in $n(\mu,T)$ 
 when one considers $\Cr(n,T)$
as in Fig.~\ref{fig:C}a.


\begin{figure}
\includegraphics[width=0.6\columnwidth]{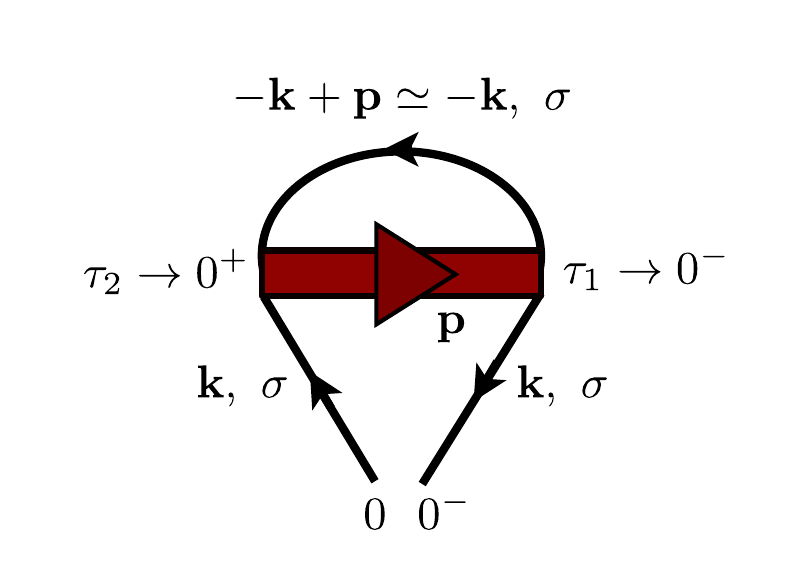}
\caption{
Leading diagrammatic contribution to the momentum distribution $n_\sigma(\kk)$ at large $k$,
which can be interpreted physically 
as the simultaneous propagation of two opposite-spin particles of large and nearly opposite momenta 
and of a missing pair  of lower momentum $p\ll k$.
Imaginary time runs from right to left.
The single-particle lines propagate forward in time 
and can be replaced by vacuum propagators. 
The pair propagator runs backwards in time and is fully dressed.
\label{fig:diag}}
\end{figure}

We turn to the momentum distribution,
and begin with an analytic observation.
The tail of the momentum distribution comes exclusively from the diagram of Fig.~\ref{fig:diag}.
Contributions from higher-order diagrams are suppressed, because integrations over internal times are restricted to narrow ranges, $G$ and $\Gamma$ being narrow functions of time at large momentum.
The corresponding asymptotic behavior of the self-energy is~\footnote{
See Ref.~\cite{BDMC_long} for detailed derivations, and Refs.~\cite{Hu_C,CombescotC} for
 related discussions restricted to the framework of $T$-matrix approximations.}
\footnote{Equation~(\ref{eq:Sig_an}) can be rederived starting from Eq.~(3.36) of~\cite{NishidaHardProbes} (Y. Nishida, {\it private communication}).}
\be
\Sigma(\kk,\tau)
\simeq
\Cr\,e^{\epsilon_k\tau},
\ \ \ \ k\to\infty, \; \tau\to0^-.
\label{eq:Sig_an}
\ee

This analytical understanding is readily incorporated into our BDMC scheme.
The $\Cr/k^4$ tail of the momentum distribution
is automatically built in provided
we
evaluate the lowest order self-energy diagram with high precision.
To do so,
we do not use Monte Carlo sampling,
 but rather the numerical procedure of Ref.~\cite{Haussmann_PRB}, the 
only essential difference being that in our case, the
pair propagator $\Gamma$ which enters the numerical procedure is the fully dressed one.

\begin{figure}
\includegraphics[width=\columnwidth,clip=,trim={0.6cm 0.5cm 1.5cm 2.5cm}]{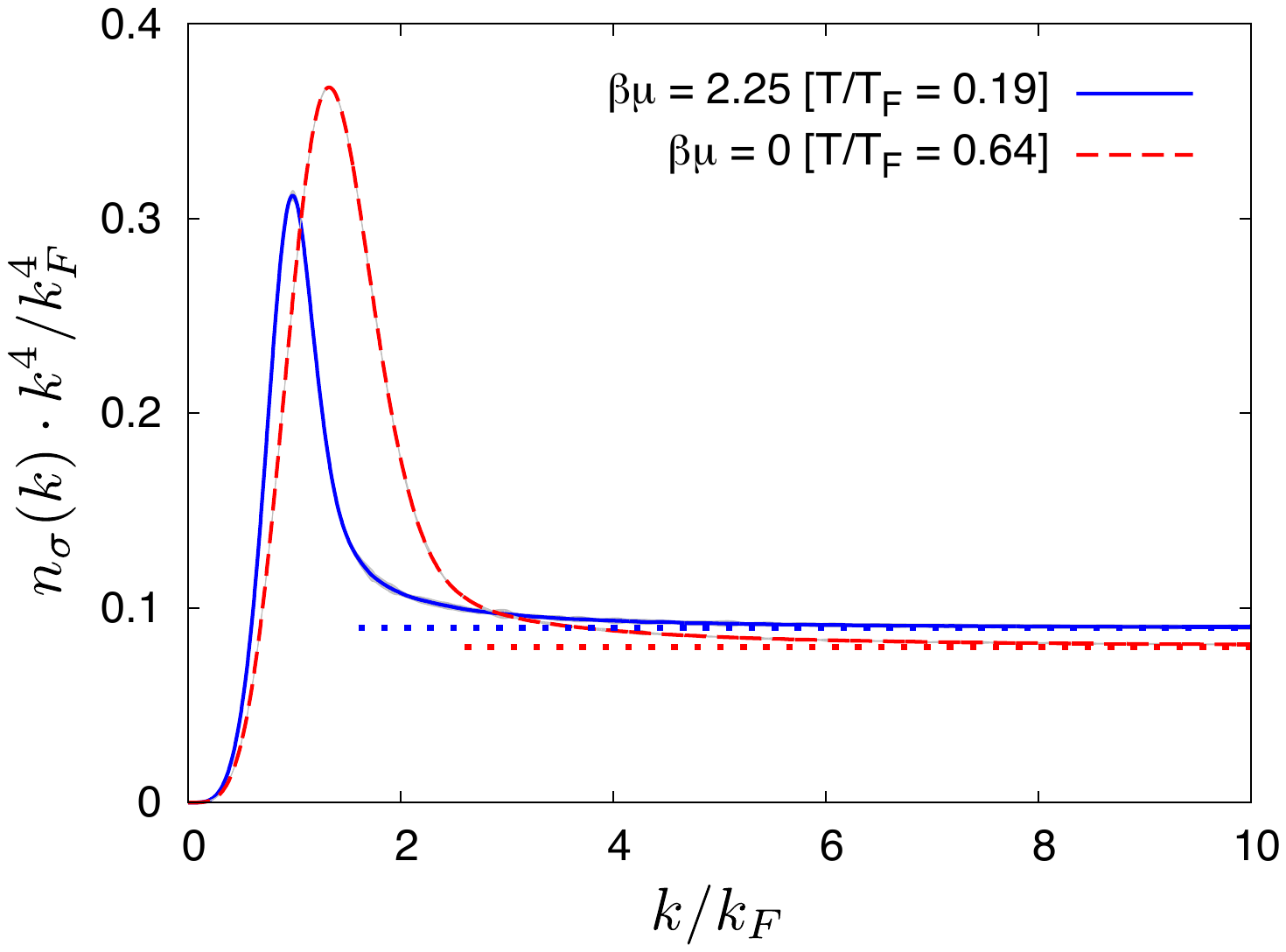}
\caption{BDMC data for the momentum distribution $n_\sigma(k)$, multiplied by $k^4$ in order to reveal the large-momentum tail
$n_\sigma(k) \sim \Cr / k^4$.
Dotted horizontal lines: values of the contact $\Cr$ computed directly from the pair propagator.
The uncertainties are represented by the grey error bands.
\label{fig:nk_k4}}
\end{figure}

\begin{figure}
\includegraphics[width=\columnwidth,clip=,trim={0.6cm 0.4cm 1.5cm 2.5cm}]{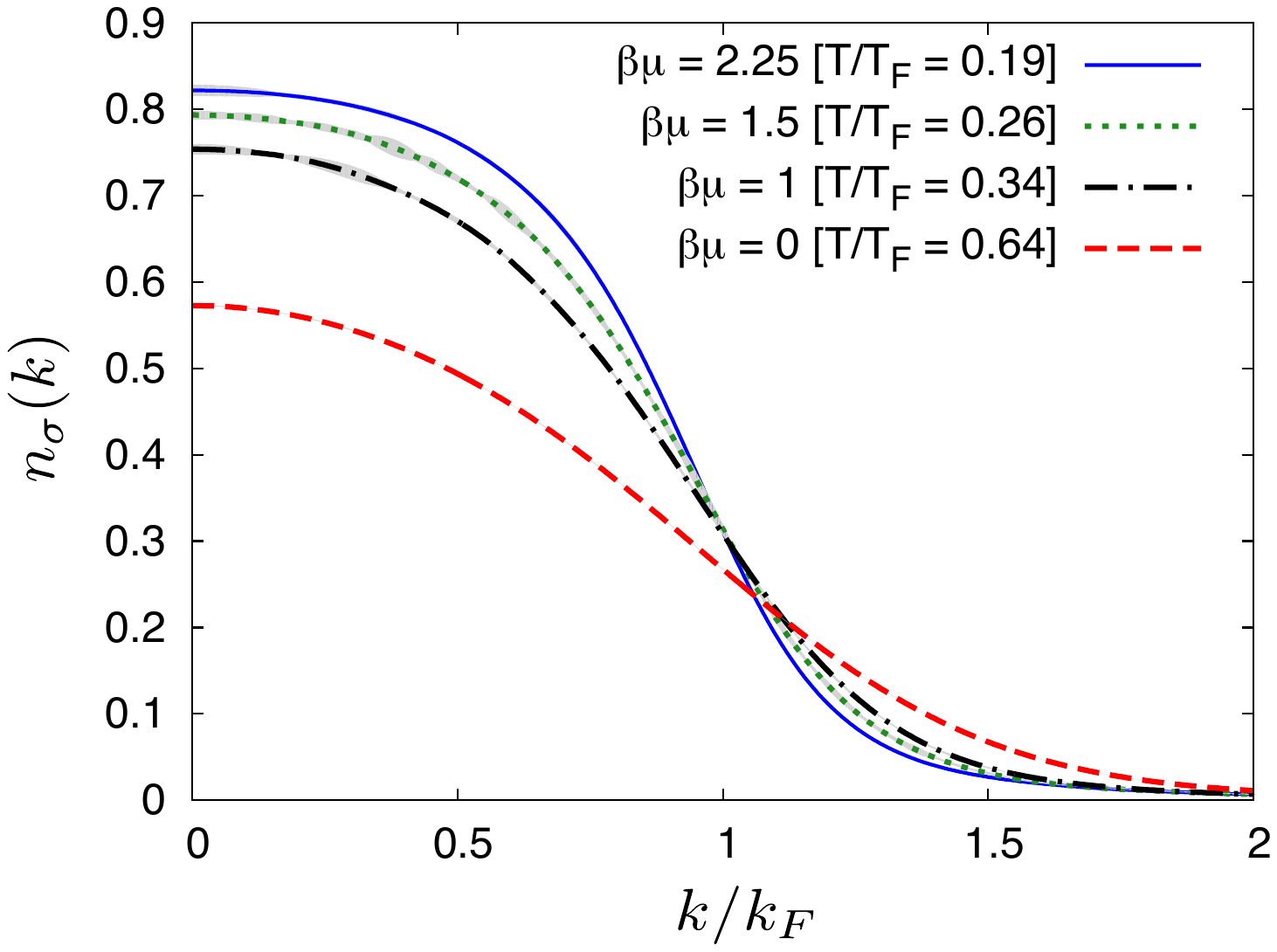}
\caption{BDMC data for the momentum distribution at various temperatures.
Error bars are represented by the grey error bands.
\label{fig:nk}}
\end{figure}

\begin{figure}
\includegraphics[width=\columnwidth,trim={0cm 2cm 0cm 1cm}]{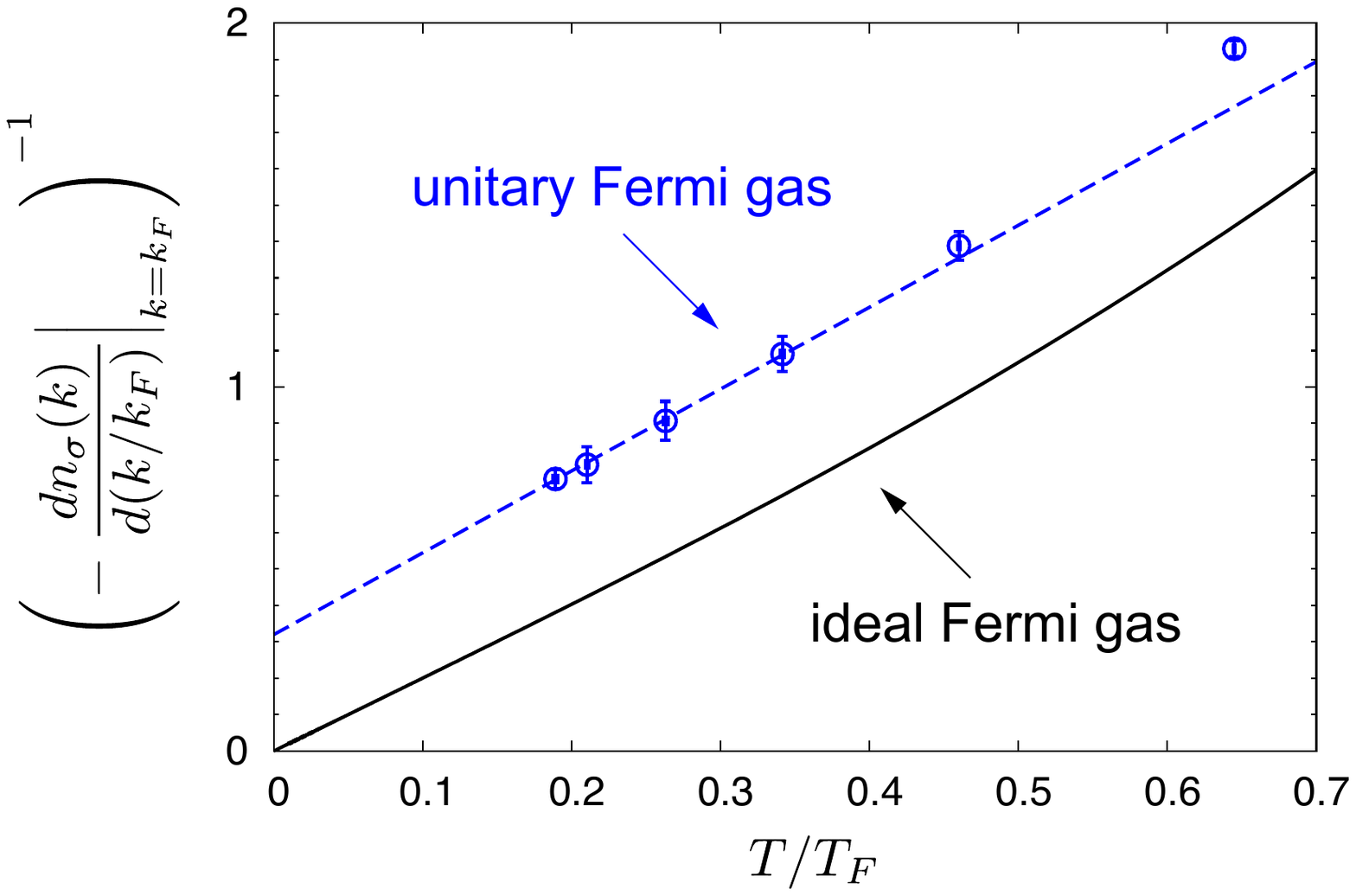}
\caption{Inverse slope of the momentum distribution at the Fermi momentum {\it vs.} temperature. 
For a Fermi liquid this quantity linearly tends to zero for $T/T_F\to0$ (see solid line). In contrast, a linear extrapolation of our data for the unitary Fermi gas (dashed line) does not go through the origin. 
\label{fig:slope}}
\end{figure}

The momentum distribution times $k^4$
is shown in Fig.~\ref{fig:nk_k4} for 
two different temperatures.
The large-momentum tail is reproduced 
without $k$-dependent statistical noise---in sharp contrast to other Monte Carlo methods~\cite{Drut_C,Carlson_C}---and perfectly agrees with our value of the contact determined from Eq.~(\ref{eq:CvsGamma}).
One can note that
the $\Cr/k^4$  tail contains
as much as $10$ to $15$ percent of the particles.
Finally, the momentum distribution at four different temperatures is shown in Fig.~\ref{fig:nk}.
The temperature dependence of $n_\sigma(k)$ is rather weak for the lowest three temperatures,
but there is no sharp feature around $k_F$
as would be the case for a pronounced degenerate Fermi liquid behavior.
The smoothness of $n_\sigma(k)$ cannot be explained by finite-temperature Fermi-liquid theory: In a Fermi liquid the slope $dn_\sigma(k)/d(k/k_F)|_{k=k_F}$ would extrapolate to $-\infty$ in the limit $T/T_F\to0$, and this does not occur for the unitary Fermi gas as shown by our data in Fig.~\ref{fig:slope}.
Deviations from Fermi liquid theory are also present in the equation of state,
{\it e.g.} the specific heat is not linear in temperature~\cite{KuEOS}.
Going to the largest temperature of Fig.~\ref{fig:nk},
the low-momentum occupation numbers become much more depleted and the distribution broadens.
However, the contact $\Cr/k_F^4$ is roughly unchanged (cf. Fig.~\ref{fig:C}a),
which can be viewed as a delicate compensation between two trends:
The
 occupation numbers increase  for $k$ moderately larger than $k_F$, which tends to increase the contact,
but the onset of the $\Cr/k^4$ regime is pushed to higher momenta (see Fig.~\ref{fig:nk_k4}), which tends to decrease the contact.

In conclusion, 
we obtained 
accurate
results for the temperature dependence of the contact and the momentum distribution
of the normal unitary Fermi gas.
This allows to discriminate between the contradicting earlier predictions. 
 In the canonical ensemble,
the contact is found to depend only weakly on temperature
in a broad temperature range $T \! \lesssim \! T_F$,
in remarkable agreement with
the self-consistent $T$-matrix approximation~\cite{ZwergerViscosity}. 
The experimental data~\cite{Jin_C_homogeneous}
are also 
consistent with our results 
given the experimental error bars~\footnote{
New preliminary
experimental data from MIT are also in agreement with our results (J.~Struck, {\it private communication}).}.
More accurate experimental data are highly desirable to provide a more stringent test of our theoretical approach, and
to understand the behavior of the contact when crossing the superfluid phase transition~\cite{Jin_C_homogeneous}.
Our results can also serve as benchmarks
in numerous contexts
 where the contact
appears
in sum rules~\cite{TanEnergetics,ZwergerRF,BaymRF,TaylorRanderiaViscosity}
or in
ultraviolet asymptotics
~\cite{Son_S_kw,TaylorRanderiaViscosity,Goldberger_S_kw,
ZwergerViscosity,
NishidaHardProbes,EnssHaussmann_Spin_Diff,Hofmann_OPE}.

\begin{acknowledgements}
{\it Acknowledgements.}
We thank F.~Chevy for fruitful discussions.
The data of Refs.~\cite{Jin_C_homogeneous,Drut_C,StrinatiC,Hu_C,ZwergerViscosity,ChapStrinatiBref,HaussmannZwergerThermo,Hu_PRA_2008,SunVirial3,GoulkoProceeding2010,Goulko_UFG_2016} were kindly provided by their authors.
This work was supported by
ERC grants Thermodynamix and Critisup2 (FW),
a PICS from CNRS (FW, NP and BS),
National Science Foundation under grant DMR-1720465
and MURI Program ``Advanced quantum materials -- a new frontier for ultracold atoms''
from AFOSR (NP and BS),
and the Simons Collaboration on the Many Electron Problem (EK, NP and BS).
TO was supported
by the MEXT HPCI
Strategic Programs for Innovative Research (SPIRE),
the Computational Materials Science Initiative (CMSI)
and Creation of New Functional Devices and High-Performance Materials to Support Next Generation Industries (CDMSI), and by
a Grant-in-Aid for Scientific Research 
(No.  22104010,
 22340090, 16H06345 and 18K13477)
from MEXT, Japan.
Simulations ran on clusters
at
LKB-LPTMC/UPMC, 
UMass,  
the Supercomputer
Center of the Institute for Solid State Physics at the University of Tokyo,
and on the K computer provided by the RIKEN Advanced Institute for
Computational Science under the HPCI System Research project (project number hp130007, hp140215,
hp150211, hp160201, and hp170263).
We acknowledge the hospitality of
the Institute for Nuclear Theory, Seattle,
the Aspen Center for Physics,
and Mainz Institute for Theoretical Physics.

\end{acknowledgements}

\bibliography{felix_copy}

\begin{thebibliography}{59}
\expandafter\ifx\csname natexlab\endcsname\relax\def\natexlab#1{#1}\fi
\expandafter\ifx\csname bibnamefont\endcsname\relax
  \def\bibnamefont#1{#1}\fi
\expandafter\ifx\csname bibfnamefont\endcsname\relax
  \def\bibfnamefont#1{#1}\fi
\expandafter\ifx\csname citenamefont\endcsname\relax
  \def\citenamefont#1{#1}\fi
\expandafter\ifx\csname url\endcsname\relax
  \def\url#1{\texttt{#1}}\fi
\expandafter\ifx\csname urlprefix\endcsname\relax\def\urlprefix{URL }\fi
\providecommand{\bibinfo}[2]{#2}
\providecommand{\eprint}[2][]{\url{#2}}

\bibitem[{\citenamefont{Leggett}()}]{Leggett1980_1}
\bibinfo{author}{\bibfnamefont{A.~J.} \bibnamefont{Leggett}}, \eprint{in: A.
  Pekalski, J. Przystawa (eds.), {\sl Modern Trends in the Theory of Condensed
  Matter}, p. 13. Springer, New York (1980)}.

\bibitem[{\citenamefont{Leggett}(1980)}]{Leggett1980_2}
\bibinfo{author}{\bibfnamefont{A.~J.} \bibnamefont{Leggett}},
  \bibinfo{journal}{J. Phys. (Paris)} \textbf{\bibinfo{volume}{42}},
  \bibinfo{pages}{C7} (\bibinfo{year}{1980}).

\bibitem[{\citenamefont{Nozi{\`e}res and Schmitt-Rink}(1985)}]{NSR}
\bibinfo{author}{\bibfnamefont{P.}~\bibnamefont{Nozi{\`e}res}}
  \bibnamefont{and}
  \bibinfo{author}{\bibfnamefont{S.}~\bibnamefont{Schmitt-Rink}},
  \bibinfo{journal}{J. Low Temp. Phys.} \textbf{\bibinfo{volume}{59}},
  \bibinfo{pages}{195} (\bibinfo{year}{1985}).

\bibitem[{\citenamefont{Leggett and Zhang}(2012)}]{ChapLeggettBref}
\bibinfo{author}{\bibfnamefont{A.~J.} \bibnamefont{Leggett}} \bibnamefont{and}
  \bibinfo{author}{\bibfnamefont{S.}~\bibnamefont{Zhang}},
  \bibinfo{journal}{Lecture Notes in Physics} \textbf{\bibinfo{volume}{836}},
  \bibinfo{pages}{33} (\bibinfo{year}{2012}), \bibinfo{note}{in
  \cite{ZwergerBook}}.

\bibitem[{Zwe()}]{ZwergerBook}
\eprint{{\sl The BCS-BEC Crossover and the Unitary Fermi Gas}, Lecture Notes in
  Physics {\bf 836}, W. Zwerger ed. (Springer, Heidelberg, 2012)}.

\bibitem[{\citenamefont{Carlson et~al.}(2012)\citenamefont{Carlson, Gandolfi,
  and Gezerlis}}]{GezerlisNeutronsReview}
\bibinfo{author}{\bibfnamefont{J.}~\bibnamefont{Carlson}},
  \bibinfo{author}{\bibfnamefont{S.}~\bibnamefont{Gandolfi}}, \bibnamefont{and}
  \bibinfo{author}{\bibfnamefont{A.}~\bibnamefont{Gezerlis}},
  \bibinfo{journal}{Prog. Theor. Exp. Phys.} \textbf{\bibinfo{volume}{2012}},
  \bibinfo{pages}{01A209} (\bibinfo{year}{2012}).

\bibitem[{\citenamefont{Strinati et~al.}(2018)\citenamefont{Strinati, Pieri,
  Roepke, Schuck, and Urban}}]{StrinatiUrbanReview}
\bibinfo{author}{\bibfnamefont{G.~C.} \bibnamefont{Strinati}},
  \bibinfo{author}{\bibfnamefont{P.}~\bibnamefont{Pieri}},
  \bibinfo{author}{\bibfnamefont{G.}~\bibnamefont{Roepke}},
  \bibinfo{author}{\bibfnamefont{P.}~\bibnamefont{Schuck}}, \bibnamefont{and}
  \bibinfo{author}{\bibfnamefont{M.}~\bibnamefont{Urban}},
  \bibinfo{journal}{Phys. Rep.} \textbf{\bibinfo{volume}{738}},
  \bibinfo{pages}{1} (\bibinfo{year}{2018}).

\bibitem[{NJP()}]{NJP_QCD}
\eprint{{New J. Phys {\bf 14} (2011), {\sl Focus on Strongly Correlated Quantum
  Fluids: from Ultracold Quantum Gases to QCD Plasmas}, A. Adams, L. D. Carr,
  T. Schaefer, P. Steinberg, J. E. Thomas (eds.)}}.

\bibitem[{\citenamefont{Tan}(2008{\natexlab{a}})}]{TanEnergetics}
\bibinfo{author}{\bibfnamefont{S.}~\bibnamefont{Tan}}, \bibinfo{journal}{Ann.
  Phys.} \textbf{\bibinfo{volume}{323}}, \bibinfo{pages}{2952}
  (\bibinfo{year}{2008}{\natexlab{a}}).

\bibitem[{\citenamefont{Tan}(2008{\natexlab{b}})}]{TanLargeMomentum}
\bibinfo{author}{\bibfnamefont{S.}~\bibnamefont{Tan}}, \bibinfo{journal}{Ann.
  Phys.} \textbf{\bibinfo{volume}{323}}, \bibinfo{pages}{2971}
  (\bibinfo{year}{2008}{\natexlab{b}}).

\bibitem[{\citenamefont{Braaten}(2012)}]{ChapBraatenBref}
\bibinfo{author}{\bibfnamefont{E.}~\bibnamefont{Braaten}},
  \bibinfo{journal}{Lecture Notes in Physics} \textbf{\bibinfo{volume}{836}},
  \bibinfo{pages}{193} (\bibinfo{year}{2012}), \bibinfo{note}{in
  \cite{ZwergerBook}}.

\bibitem[{\citenamefont{Castin and Werner}(2012)}]{LeChapitreIn}
\bibinfo{author}{\bibfnamefont{Y.}~\bibnamefont{Castin}} \bibnamefont{and}
  \bibinfo{author}{\bibfnamefont{F.}~\bibnamefont{Werner}},
  \bibinfo{journal}{Lecture Notes in Physics} \textbf{\bibinfo{volume}{836}},
  \bibinfo{pages}{127} (\bibinfo{year}{2012}),
  \bibinfo{note}{in~\cite{ZwergerBook}}.

\bibitem[{\citenamefont{Braaten and Platter}(2008)}]{BraatenC}
\bibinfo{author}{\bibfnamefont{E.}~\bibnamefont{Braaten}} \bibnamefont{and}
  \bibinfo{author}{\bibfnamefont{L.}~\bibnamefont{Platter}},
  \bibinfo{journal}{Phys. Rev. Lett.} \textbf{\bibinfo{volume}{100}},
  \bibinfo{pages}{205301} (\bibinfo{year}{2008}).

\bibitem[{\citenamefont{Partridge et~al.}(2005)\citenamefont{Partridge,
  Strecker, Kamar, Jack, and Hulet}}]{HuletClosedChannel}
\bibinfo{author}{\bibfnamefont{G.~B.} \bibnamefont{Partridge}},
  \bibinfo{author}{\bibfnamefont{K.~E.} \bibnamefont{Strecker}},
  \bibinfo{author}{\bibfnamefont{R.~I.} \bibnamefont{Kamar}},
  \bibinfo{author}{\bibfnamefont{M.~W.} \bibnamefont{Jack}}, \bibnamefont{and}
  \bibinfo{author}{\bibfnamefont{R.~G.} \bibnamefont{Hulet}},
  \bibinfo{journal}{Phys. Rev. Lett.} \textbf{\bibinfo{volume}{95}},
  \bibinfo{eid}{020404} (\bibinfo{year}{2005}).

\bibitem[{\citenamefont{Werner et~al.}(2009)\citenamefont{Werner, Tarruell, and
  Castin}}]{WTC}
\bibinfo{author}{\bibfnamefont{F.}~\bibnamefont{Werner}},
  \bibinfo{author}{\bibfnamefont{L.}~\bibnamefont{Tarruell}}, \bibnamefont{and}
  \bibinfo{author}{\bibfnamefont{Y.}~\bibnamefont{Castin}},
  \bibinfo{journal}{Eur. Phys. J. B} \textbf{\bibinfo{volume}{68}},
  \bibinfo{pages}{401} (\bibinfo{year}{2009}).

\bibitem[{\citenamefont{Kuhnle et~al.}(2010)\citenamefont{Kuhnle, Hu, Liu,
  Dyke, Mark, Drummond, Hannaford, and Vale}}]{AustraliensC}
\bibinfo{author}{\bibfnamefont{E.~D.} \bibnamefont{Kuhnle}},
  \bibinfo{author}{\bibfnamefont{H.}~\bibnamefont{Hu}},
  \bibinfo{author}{\bibfnamefont{X.-J.} \bibnamefont{Liu}},
  \bibinfo{author}{\bibfnamefont{P.}~\bibnamefont{Dyke}},
  \bibinfo{author}{\bibfnamefont{M.}~\bibnamefont{Mark}},
  \bibinfo{author}{\bibfnamefont{P.~D.} \bibnamefont{Drummond}},
  \bibinfo{author}{\bibfnamefont{P.}~\bibnamefont{Hannaford}},
  \bibnamefont{and} \bibinfo{author}{\bibfnamefont{C.~J.} \bibnamefont{Vale}},
  \bibinfo{journal}{Phys. Rev. Lett.} \textbf{\bibinfo{volume}{105}},
  \bibinfo{pages}{070402} (\bibinfo{year}{2010}).

\bibitem[{\citenamefont{Kuhnle et~al.}(2011)\citenamefont{Kuhnle, Hoinka, Dyke,
  Hu, Hannaford, and Vale}}]{AustraliensT}
\bibinfo{author}{\bibfnamefont{E.~D.} \bibnamefont{Kuhnle}},
  \bibinfo{author}{\bibfnamefont{S.}~\bibnamefont{Hoinka}},
  \bibinfo{author}{\bibfnamefont{P.}~\bibnamefont{Dyke}},
  \bibinfo{author}{\bibfnamefont{H.}~\bibnamefont{Hu}},
  \bibinfo{author}{\bibfnamefont{P.}~\bibnamefont{Hannaford}},
  \bibnamefont{and} \bibinfo{author}{\bibfnamefont{C.~J.} \bibnamefont{Vale}},
  \bibinfo{journal}{Phys. Rev. Lett.} \textbf{\bibinfo{volume}{106}},
  \bibinfo{pages}{170402} (\bibinfo{year}{2011}).

\bibitem[{\citenamefont{Hoinka et~al.}(2013)\citenamefont{Hoinka, Lingham,
  Fenech, Hu, Vale, Drut, and Gandolfi}}]{ValeC_precise}
\bibinfo{author}{\bibfnamefont{S.}~\bibnamefont{Hoinka}},
  \bibinfo{author}{\bibfnamefont{M.}~\bibnamefont{Lingham}},
  \bibinfo{author}{\bibfnamefont{K.}~\bibnamefont{Fenech}},
  \bibinfo{author}{\bibfnamefont{H.}~\bibnamefont{Hu}},
  \bibinfo{author}{\bibfnamefont{C.~J.} \bibnamefont{Vale}},
  \bibinfo{author}{\bibfnamefont{J.~E.} \bibnamefont{Drut}}, \bibnamefont{and}
  \bibinfo{author}{\bibfnamefont{S.}~\bibnamefont{Gandolfi}},
  \bibinfo{journal}{Phys. Rev. Lett.} \textbf{\bibinfo{volume}{110}},
  \bibinfo{pages}{055305} (\bibinfo{year}{2013}).

\bibitem[{\citenamefont{Stewart et~al.}(2010)\citenamefont{Stewart, Gaebler,
  Drake, and Jin}}]{JinUnivRel}
\bibinfo{author}{\bibfnamefont{J.~T.} \bibnamefont{Stewart}},
  \bibinfo{author}{\bibfnamefont{J.~P.} \bibnamefont{Gaebler}},
  \bibinfo{author}{\bibfnamefont{T.~E.} \bibnamefont{Drake}}, \bibnamefont{and}
  \bibinfo{author}{\bibfnamefont{D.~S.} \bibnamefont{Jin}},
  \bibinfo{journal}{Phys. Rev. Lett.} \textbf{\bibinfo{volume}{104}},
  \bibinfo{pages}{235301} (\bibinfo{year}{2010}).

\bibitem[{\citenamefont{Navon et~al.}(2010)\citenamefont{Navon, Nascimb{\`e}ne,
  Chevy, and Salomon}}]{NirEOS}
\bibinfo{author}{\bibfnamefont{N.}~\bibnamefont{Navon}},
  \bibinfo{author}{\bibfnamefont{S.}~\bibnamefont{Nascimb{\`e}ne}},
  \bibinfo{author}{\bibfnamefont{F.}~\bibnamefont{Chevy}}, \bibnamefont{and}
  \bibinfo{author}{\bibfnamefont{C.}~\bibnamefont{Salomon}},
  \bibinfo{journal}{Science} \textbf{\bibinfo{volume}{328}},
  \bibinfo{pages}{729} (\bibinfo{year}{2010}).

\bibitem[{\citenamefont{Schneider and Randeria}(2010)}]{RanderiaRF}
\bibinfo{author}{\bibfnamefont{W.}~\bibnamefont{Schneider}} \bibnamefont{and}
  \bibinfo{author}{\bibfnamefont{M.}~\bibnamefont{Randeria}},
  \bibinfo{journal}{Phys. Rev. A} \textbf{\bibinfo{volume}{81}},
  \bibinfo{pages}{021601(R)} (\bibinfo{year}{2010}).

\bibitem[{\citenamefont{Sagi et~al.}(2012)\citenamefont{Sagi, Drake, Paudel,
  and Jin}}]{Jin_C_homogeneous}
\bibinfo{author}{\bibfnamefont{Y.}~\bibnamefont{Sagi}},
  \bibinfo{author}{\bibfnamefont{T.~E.} \bibnamefont{Drake}},
  \bibinfo{author}{\bibfnamefont{R.}~\bibnamefont{Paudel}}, \bibnamefont{and}
  \bibinfo{author}{\bibfnamefont{D.~S.} \bibnamefont{Jin}},
  \bibinfo{journal}{Phys. Rev. Lett.} \textbf{\bibinfo{volume}{109}},
  \bibinfo{pages}{220402} (\bibinfo{year}{2012}).

\bibitem[{\citenamefont{{Shkedrov} et~al.}()\citenamefont{{Shkedrov},
  {Florshaim}, {Ness}, {Gandman}, and {Sagi}}}]{SagiContactRF}
\bibinfo{author}{\bibfnamefont{C.}~\bibnamefont{{Shkedrov}}},
  \bibinfo{author}{\bibfnamefont{Y.}~\bibnamefont{{Florshaim}}},
  \bibinfo{author}{\bibfnamefont{G.}~\bibnamefont{{Ness}}},
  \bibinfo{author}{\bibfnamefont{A.}~\bibnamefont{{Gandman}}},
  \bibnamefont{and} \bibinfo{author}{\bibfnamefont{Y.}~\bibnamefont{{Sagi}}},
  \emph{\bibinfo{title}{{High Sensitivity RF Spectroscopy of a
  Strongly-Interacting Fermi Gas}}}, \eprint{arXiv:1803.01770}.

\bibitem[{\citenamefont{Laurent et~al.}(2017)\citenamefont{Laurent, Pierce,
  Delehaye, Yefsah, Chevy, and Salomon}}]{LaurentC}
\bibinfo{author}{\bibfnamefont{S.}~\bibnamefont{Laurent}},
  \bibinfo{author}{\bibfnamefont{M.}~\bibnamefont{Pierce}},
  \bibinfo{author}{\bibfnamefont{M.}~\bibnamefont{Delehaye}},
  \bibinfo{author}{\bibfnamefont{T.}~\bibnamefont{Yefsah}},
  \bibinfo{author}{\bibfnamefont{F.}~\bibnamefont{Chevy}}, \bibnamefont{and}
  \bibinfo{author}{\bibfnamefont{C.}~\bibnamefont{Salomon}},
  \bibinfo{journal}{Phys. Rev. Lett.} \textbf{\bibinfo{volume}{118}},
  \bibinfo{pages}{103403} (\bibinfo{year}{2017}).

\bibitem[{Con()}]{ContactSantFeliu}
\eprint{Talk by C. Vale (Swinburne University) and poster by J.~Struck (MIT) at
  the conference {\it BEC 2017 - Frontiers in Quantum Gases}, Sant Feliu de
  Guixols, Spain}.

\bibitem[{\citenamefont{Palestini et~al.}(2010)\citenamefont{Palestini, Perali,
  Pieri, and Strinati}}]{StrinatiC}
\bibinfo{author}{\bibfnamefont{F.}~\bibnamefont{Palestini}},
  \bibinfo{author}{\bibfnamefont{A.}~\bibnamefont{Perali}},
  \bibinfo{author}{\bibfnamefont{P.}~\bibnamefont{Pieri}}, \bibnamefont{and}
  \bibinfo{author}{\bibfnamefont{G.~C.} \bibnamefont{Strinati}},
  \bibinfo{journal}{Phys. Rev. A} \textbf{\bibinfo{volume}{82}},
  \bibinfo{pages}{021605(R)} (\bibinfo{year}{2010}).

\bibitem[{\citenamefont{Enss et~al.}(2011)\citenamefont{Enss, Haussmann, and
  Zwerger}}]{ZwergerViscosity}
\bibinfo{author}{\bibfnamefont{T.}~\bibnamefont{Enss}},
  \bibinfo{author}{\bibfnamefont{R.}~\bibnamefont{Haussmann}},
  \bibnamefont{and} \bibinfo{author}{\bibfnamefont{W.}~\bibnamefont{Zwerger}},
  \bibinfo{journal}{Ann. Phys.} \textbf{\bibinfo{volume}{326}},
  \bibinfo{pages}{770} (\bibinfo{year}{2011}).

\bibitem[{\citenamefont{Hu et~al.}(2011)\citenamefont{Hu, Liu, and
  Drummond}}]{Hu_C}
\bibinfo{author}{\bibfnamefont{H.}~\bibnamefont{Hu}},
  \bibinfo{author}{\bibfnamefont{X.-J.} \bibnamefont{Liu}}, \bibnamefont{and}
  \bibinfo{author}{\bibfnamefont{P.~D.} \bibnamefont{Drummond}},
  \bibinfo{journal}{New J. Phys.} \textbf{\bibinfo{volume}{13}},
  \bibinfo{pages}{035007} (\bibinfo{year}{2011}).

\bibitem[{\citenamefont{Drut et~al.}(2011)\citenamefont{Drut, L{\"a}hde, and
  Ten}}]{Drut_C}
\bibinfo{author}{\bibfnamefont{J.~E.} \bibnamefont{Drut}},
  \bibinfo{author}{\bibfnamefont{T.~A.} \bibnamefont{L{\"a}hde}},
  \bibnamefont{and} \bibinfo{author}{\bibfnamefont{T.}~\bibnamefont{Ten}},
  \bibinfo{journal}{Phys. Rev. Lett.} \textbf{\bibinfo{volume}{106}},
  \bibinfo{pages}{205302} (\bibinfo{year}{2011}).

\bibitem[{\citenamefont{Goulko and Wingate}(2016)}]{Goulko_UFG_2016}
\bibinfo{author}{\bibfnamefont{O.}~\bibnamefont{Goulko}} \bibnamefont{and}
  \bibinfo{author}{\bibfnamefont{M.}~\bibnamefont{Wingate}},
  \bibinfo{journal}{Phys. Rev. A} \textbf{\bibinfo{volume}{93}},
  \bibinfo{pages}{053604} (\bibinfo{year}{2016}).

\bibitem[{\citenamefont{{Van Houcke} et~al.}(2012)\citenamefont{{Van Houcke},
  Werner, Kozik, Prokof'ev, Svistunov, Ku, Sommer, Cheuk, Schirotzek, and
  Zwierlein}}]{VanHouckeEOS}
\bibinfo{author}{\bibfnamefont{K.}~\bibnamefont{{Van Houcke}}},
  \bibinfo{author}{\bibfnamefont{F.}~\bibnamefont{Werner}},
  \bibinfo{author}{\bibfnamefont{E.}~\bibnamefont{Kozik}},
  \bibinfo{author}{\bibfnamefont{N.}~\bibnamefont{Prokof'ev}},
  \bibinfo{author}{\bibfnamefont{B.}~\bibnamefont{Svistunov}},
  \bibinfo{author}{\bibfnamefont{M.~J.~H.} \bibnamefont{Ku}},
  \bibinfo{author}{\bibfnamefont{A.~T.} \bibnamefont{Sommer}},
  \bibinfo{author}{\bibfnamefont{L.~W.} \bibnamefont{Cheuk}},
  \bibinfo{author}{\bibfnamefont{A.}~\bibnamefont{Schirotzek}},
  \bibnamefont{and} \bibinfo{author}{\bibfnamefont{M.~W.}
  \bibnamefont{Zwierlein}}, \bibinfo{journal}{Nature Phys.}
  \textbf{\bibinfo{volume}{8}}, \bibinfo{pages}{366} (\bibinfo{year}{2012}).

\bibitem[{\citenamefont{Rossi et~al.}({\natexlab{a}})\citenamefont{Rossi,
  Ohgoe, {Van Houcke}, and Werner}}]{RossiEOS}
\bibinfo{author}{\bibfnamefont{R.}~\bibnamefont{Rossi}},
  \bibinfo{author}{\bibfnamefont{T.}~\bibnamefont{Ohgoe}},
  \bibinfo{author}{\bibfnamefont{K.}~\bibnamefont{{Van Houcke}}},
  \bibnamefont{and} \bibinfo{author}{\bibfnamefont{F.}~\bibnamefont{Werner}},
  \emph{\bibinfo{title}{Resummation of diagrammatic series with zero
  convergence radius for strongly correlated fermions}},
  \eprint{arXiv:1802.07717}.

\bibitem[{\citenamefont{Pieri et~al.}(2009)\citenamefont{Pieri, Perali, and
  Strinati}}]{StrinatiRF}
\bibinfo{author}{\bibfnamefont{P.}~\bibnamefont{Pieri}},
  \bibinfo{author}{\bibfnamefont{A.}~\bibnamefont{Perali}}, \bibnamefont{and}
  \bibinfo{author}{\bibfnamefont{G.~C.} \bibnamefont{Strinati}},
  \bibinfo{journal}{Nature Physics} \textbf{\bibinfo{volume}{5}},
  \bibinfo{pages}{736} (\bibinfo{year}{2009}).

\bibitem[{\citenamefont{Haussmann et~al.}(2009)\citenamefont{Haussmann, Punk,
  and Zwerger}}]{ZwergerRFLong}
\bibinfo{author}{\bibfnamefont{R.}~\bibnamefont{Haussmann}},
  \bibinfo{author}{\bibfnamefont{M.}~\bibnamefont{Punk}}, \bibnamefont{and}
  \bibinfo{author}{\bibfnamefont{W.}~\bibnamefont{Zwerger}},
  \bibinfo{journal}{Phys. Rev. A} \textbf{\bibinfo{volume}{80}},
  \bibinfo{pages}{063612} (\bibinfo{year}{2009}).

\bibitem[{\citenamefont{Werner and
  Castin}(2012)}]{WernerCastinRelationsFermions}
\bibinfo{author}{\bibfnamefont{F.}~\bibnamefont{Werner}} \bibnamefont{and}
  \bibinfo{author}{\bibfnamefont{Y.}~\bibnamefont{Castin}},
  \bibinfo{journal}{Phys. Rev. A} \textbf{\bibinfo{volume}{86}},
  \bibinfo{pages}{013626} (\bibinfo{year}{2012}).

\bibitem[{\citenamefont{{Van Houcke} et~al.}()\citenamefont{{Van Houcke},
  Werner, Prokof'ev, and Svistunov}}]{BDMC_long}
\bibinfo{author}{\bibfnamefont{K.}~\bibnamefont{{Van Houcke}}},
  \bibinfo{author}{\bibfnamefont{F.}~\bibnamefont{Werner}},
  \bibinfo{author}{\bibfnamefont{N.}~\bibnamefont{Prokof'ev}},
  \bibnamefont{and}
  \bibinfo{author}{\bibfnamefont{B.}~\bibnamefont{Svistunov}},
  \emph{\bibinfo{title}{{Bold diagrammatic Monte Carlo for the resonant Fermi
  gas}}}, \eprint{arXiv:1305.3901}.

\bibitem[{\citenamefont{Rossi et~al.}({\natexlab{b}})\citenamefont{Rossi,
  Ohgoe, {Van Houcke}, and Werner}}]{ResonLong2}
\bibinfo{author}{\bibfnamefont{R.}~\bibnamefont{Rossi}},
  \bibinfo{author}{\bibfnamefont{T.}~\bibnamefont{Ohgoe}},
  \bibinfo{author}{\bibfnamefont{K.}~\bibnamefont{{Van Houcke}}},
  \bibnamefont{and} \bibinfo{author}{\bibfnamefont{F.}~\bibnamefont{Werner}},
  \bibinfo{note}{in preparation}.

\bibitem[{\citenamefont{Haussmann}(1994)}]{Haussmann_PRB}
\bibinfo{author}{\bibfnamefont{R.}~\bibnamefont{Haussmann}},
  \bibinfo{journal}{Phys. Rev. B} \textbf{\bibinfo{volume}{49}},
  \bibinfo{pages}{12975} (\bibinfo{year}{1994}).

\bibitem[{\citenamefont{Ku et~al.}(2012)\citenamefont{Ku, Sommer, Cheuk, and
  Zwierlein}}]{KuEOS}
\bibinfo{author}{\bibfnamefont{M.~J.~H.} \bibnamefont{Ku}},
  \bibinfo{author}{\bibfnamefont{A.}~\bibnamefont{Sommer}},
  \bibinfo{author}{\bibfnamefont{L.~W.} \bibnamefont{Cheuk}}, \bibnamefont{and}
  \bibinfo{author}{\bibfnamefont{M.~W.} \bibnamefont{Zwierlein}},
  \bibinfo{journal}{Science} \textbf{\bibinfo{volume}{335}},
  \bibinfo{pages}{563} (\bibinfo{year}{2012}).

\bibitem[{\citenamefont{Goulko and Wingate}(2006)}]{GoulkoBurovski}
\bibinfo{author}{\bibfnamefont{O.}~\bibnamefont{Goulko}} \bibnamefont{and}
  \bibinfo{author}{\bibfnamefont{M.}~\bibnamefont{Wingate}},
  \bibinfo{journal}{Phys. Rev. A {\bf 82}, 053621 (2010); E. Burovski, N.
  Prokof’ev, B. Svistunov, M. Troyer, Phys. Rev. Lett. {\bf 96}, 160402}
  (\bibinfo{year}{2006}).

\bibitem[{\citenamefont{Yu et~al.}(2009)\citenamefont{Yu, Bruun, and
  Baym}}]{Baym_C}
\bibinfo{author}{\bibfnamefont{Z.}~\bibnamefont{Yu}},
  \bibinfo{author}{\bibfnamefont{G.~M.} \bibnamefont{Bruun}}, \bibnamefont{and}
  \bibinfo{author}{\bibfnamefont{G.}~\bibnamefont{Baym}},
  \bibinfo{journal}{Phys. Rev. A} \textbf{\bibinfo{volume}{80}},
  \bibinfo{pages}{023615} (\bibinfo{year}{2009}).

\bibitem[{\citenamefont{Sun and Leyronas}(2015)}]{SunVirial3}
\bibinfo{author}{\bibfnamefont{M.}~\bibnamefont{Sun}} \bibnamefont{and}
  \bibinfo{author}{\bibfnamefont{X.}~\bibnamefont{Leyronas}},
  \bibinfo{journal}{Phys. Rev. A} \textbf{\bibinfo{volume}{92}},
  \bibinfo{pages}{053611} (\bibinfo{year}{2015}).

\bibitem[{\citenamefont{Leyronas}()}]{LeyronasPrivate}
\bibinfo{author}{\bibfnamefont{X.}~\bibnamefont{Leyronas}}, \bibinfo{note}{{\it
  private communication}}.

\bibitem[{\citenamefont{Goulko and Wingate}(2010)}]{GoulkoProceeding2010}
\bibinfo{author}{\bibfnamefont{O.}~\bibnamefont{Goulko}} \bibnamefont{and}
  \bibinfo{author}{\bibfnamefont{M.}~\bibnamefont{Wingate}},
  \bibinfo{journal}{PoS Lattice2010} \textbf{\bibinfo{volume}{187}}
  (\bibinfo{year}{2010}).

\bibitem[{\citenamefont{Strinati}(2012)}]{ChapStrinatiBref}
\bibinfo{author}{\bibfnamefont{G.~C.} \bibnamefont{Strinati}},
  \bibinfo{journal}{Lecture Notes in Physics} \textbf{\bibinfo{volume}{836}},
  \bibinfo{pages}{99} (\bibinfo{year}{2012}).

\bibitem[{\citenamefont{Haussmann et~al.}(2007)\citenamefont{Haussmann,
  Rantner, Cerrito, and Zwerger}}]{HaussmannZwergerThermo}
\bibinfo{author}{\bibfnamefont{R.}~\bibnamefont{Haussmann}},
  \bibinfo{author}{\bibfnamefont{W.}~\bibnamefont{Rantner}},
  \bibinfo{author}{\bibfnamefont{S.}~\bibnamefont{Cerrito}}, \bibnamefont{and}
  \bibinfo{author}{\bibfnamefont{W.}~\bibnamefont{Zwerger}},
  \bibinfo{journal}{Phys. Rev. A} \textbf{\bibinfo{volume}{75}},
  \bibinfo{pages}{023610} (\bibinfo{year}{2007}).

\bibitem[{\citenamefont{Hu et~al.}(2008)\citenamefont{Hu, Liu, and
  Drummond}}]{Hu_PRA_2008}
\bibinfo{author}{\bibfnamefont{H.}~\bibnamefont{Hu}},
  \bibinfo{author}{\bibfnamefont{X.-J.} \bibnamefont{Liu}}, \bibnamefont{and}
  \bibinfo{author}{\bibfnamefont{P.~D.} \bibnamefont{Drummond}},
  \bibinfo{journal}{Phys. Rev. A} \textbf{\bibinfo{volume}{77}},
  \bibinfo{pages}{061605(R)} (\bibinfo{year}{2008}).

\bibitem[{\citenamefont{Gandolfi et~al.}(2011)\citenamefont{Gandolfi, Schmidt,
  and Carlson}}]{Carlson_C}
\bibinfo{author}{\bibfnamefont{S.}~\bibnamefont{Gandolfi}},
  \bibinfo{author}{\bibfnamefont{K.~E.} \bibnamefont{Schmidt}},
  \bibnamefont{and} \bibinfo{author}{\bibfnamefont{J.}~\bibnamefont{Carlson}},
  \bibinfo{journal}{Phys. Rev. A} \textbf{\bibinfo{volume}{83}},
  \bibinfo{pages}{041601(R)} (\bibinfo{year}{2011}).

\bibitem[{\citenamefont{Punk and Zwerger}(2007)}]{ZwergerRF}
\bibinfo{author}{\bibfnamefont{M.}~\bibnamefont{Punk}} \bibnamefont{and}
  \bibinfo{author}{\bibfnamefont{W.}~\bibnamefont{Zwerger}},
  \bibinfo{journal}{Phys. Rev. Lett.} \textbf{\bibinfo{volume}{99}},
  \bibinfo{eid}{170404} (\bibinfo{year}{2007}).

\bibitem[{\citenamefont{Baym et~al.}(2007)\citenamefont{Baym, Pethick, Yu, and
  Zwierlein}}]{BaymRF}
\bibinfo{author}{\bibfnamefont{G.}~\bibnamefont{Baym}},
  \bibinfo{author}{\bibfnamefont{C.~J.} \bibnamefont{Pethick}},
  \bibinfo{author}{\bibfnamefont{Z.}~\bibnamefont{Yu}}, \bibnamefont{and}
  \bibinfo{author}{\bibfnamefont{M.~W.} \bibnamefont{Zwierlein}},
  \bibinfo{journal}{Phys. Rev. Lett.} \textbf{\bibinfo{volume}{99}},
  \bibinfo{pages}{190407} (\bibinfo{year}{2007}).

\bibitem[{\citenamefont{Taylor and Randeria}(2010)}]{TaylorRanderiaViscosity}
\bibinfo{author}{\bibfnamefont{E.}~\bibnamefont{Taylor}} \bibnamefont{and}
  \bibinfo{author}{\bibfnamefont{M.}~\bibnamefont{Randeria}},
  \bibinfo{journal}{Phys. Rev. A} \textbf{\bibinfo{volume}{81}},
  \bibinfo{pages}{053610} (\bibinfo{year}{2010}).

\bibitem[{\citenamefont{Son and Thompson}(2010)}]{Son_S_kw}
\bibinfo{author}{\bibfnamefont{D.~T.} \bibnamefont{Son}} \bibnamefont{and}
  \bibinfo{author}{\bibfnamefont{E.~G.} \bibnamefont{Thompson}},
  \bibinfo{journal}{Phys. Rev. A} \textbf{\bibinfo{volume}{81}},
  \bibinfo{pages}{063634} (\bibinfo{year}{2010}).

\bibitem[{\citenamefont{Goldberger and Rothstein}(2012)}]{Goldberger_S_kw}
\bibinfo{author}{\bibfnamefont{W.~D.} \bibnamefont{Goldberger}}
  \bibnamefont{and} \bibinfo{author}{\bibfnamefont{I.~Z.}
  \bibnamefont{Rothstein}}, \bibinfo{journal}{Phys. Rev. A}
  \textbf{\bibinfo{volume}{85}}, \bibinfo{pages}{013613}
  (\bibinfo{year}{2012}).

\bibitem[{\citenamefont{Nishida}(2012)}]{NishidaHardProbes}
\bibinfo{author}{\bibfnamefont{Y.}~\bibnamefont{Nishida}},
  \bibinfo{journal}{Phys. Rev. A} \textbf{\bibinfo{volume}{85}},
  \bibinfo{pages}{053643} (\bibinfo{year}{2012}).

\bibitem[{\citenamefont{Enss and Haussmann}(2012)}]{EnssHaussmann_Spin_Diff}
\bibinfo{author}{\bibfnamefont{T.}~\bibnamefont{Enss}} \bibnamefont{and}
  \bibinfo{author}{\bibfnamefont{R.}~\bibnamefont{Haussmann}},
  \bibinfo{journal}{Phys. Rev. Lett.} \textbf{\bibinfo{volume}{109}},
  \bibinfo{pages}{195303} (\bibinfo{year}{2012}).

\bibitem[{\citenamefont{Hofmann}(2011)}]{Hofmann_OPE}
\bibinfo{author}{\bibfnamefont{J.}~\bibnamefont{Hofmann}},
  \bibinfo{journal}{Phys. Rev. A} \textbf{\bibinfo{volume}{84}},
  \bibinfo{pages}{043603} (\bibinfo{year}{2011}).

\bibitem[{\citenamefont{Campostrini et~al.}(2006)\citenamefont{Campostrini,
  Hasenbusch, Pelissetto, and Vicari}}]{hasenbusch_exponents}
\bibinfo{author}{\bibfnamefont{M.}~\bibnamefont{Campostrini}},
  \bibinfo{author}{\bibfnamefont{M.}~\bibnamefont{Hasenbusch}},
  \bibinfo{author}{\bibfnamefont{A.}~\bibnamefont{Pelissetto}},
  \bibnamefont{and} \bibinfo{author}{\bibfnamefont{E.}~\bibnamefont{Vicari}},
  \bibinfo{journal}{Phys. Rev. B} \textbf{\bibinfo{volume}{74}},
  \bibinfo{pages}{144506} (\bibinfo{year}{2006}).

\bibitem[{\citenamefont{Hasenbusch}(2006)}]{hasenbusch_amplitude_ratio}
\bibinfo{author}{\bibfnamefont{M.}~\bibnamefont{Hasenbusch}},
  \bibinfo{journal}{J. Stat. Mech.} \textbf{\bibinfo{volume}{2006}},
  \bibinfo{pages}{P08019} (\bibinfo{year}{2006}).

\bibitem[{\citenamefont{Combescot et~al.}(2009)\citenamefont{Combescot,
  Alzetto, and Leyronas}}]{CombescotC}
\bibinfo{author}{\bibfnamefont{R.}~\bibnamefont{Combescot}},
  \bibinfo{author}{\bibfnamefont{F.}~\bibnamefont{Alzetto}}, \bibnamefont{and}
  \bibinfo{author}{\bibfnamefont{X.}~\bibnamefont{Leyronas}},
  \bibinfo{journal}{Phys. Rev. A} \textbf{\bibinfo{volume}{79}},
  \bibinfo{eid}{053640} (\bibinfo{year}{2009}).

\end{thebibliography}

\end{document}